\documentclass[11pt]{article}
\usepackage{jheppub} 

\usepackage[T1]{fontenc} 
\usepackage[utf8]{inputenc}
\usepackage[normalem]{ulem}

\usepackage{tikz}
\usepackage{pifont}
\usepackage{enumerate}
\usepackage[shortlabels]{enumitem}
\usepackage{amsfonts}
\usepackage{amssymb}
\usepackage{amsmath}
\usepackage{upgreek}
\usepackage{amsthm}
\usepackage{latexsym}

\usepackage{tensor}
\usepackage{xcolor}

\newcommand{\rp}{r_{+}}

\title{\textbf{Geometric and Thermodynamic Volume of Hairy Black Branes}}

\author[a,b]{Alvaro Ballon Bordo}

\affiliation[a]{Perimeter Institute, 31 Caroline Street North, Waterloo, ON, N2L 2Y5, Canada}
\affiliation[b]{Department of Physics and Astronomy, University of Waterloo,
	Waterloo, Ontario, Canada, N2L 3G1}

\emailAdd{aballonbordo@perimeterinstitute.ca}

\abstract{
With the objective to generalize previous results found for a handful of explicit solutions, we study the extended thermodynamics of a black brane with minimally coupled scalar hair in $D$-dimensional asymptotically anti-de Sitter spacetimes. Using Komar integration and the Hamiltonian formalism to calculate the conserved charges, we obtain a Smarr relation that is applicable to a wide variety of solutions and suggests a more general definition of the thermodynamic volume. This volume is found to be proportional to the geometric volume, and a simple prescription is given to calculate the constant of proportionality. Moreover, the method of Hamiltonian perturbations yields an extended first law of thermodynamics for hairy black branes, thus giving a definition for their enthalpy. These results are verified then by applying them to some of the explicit solutions that exist in the literature.

}

\begin{document}

\maketitle

\section{Introduction}

Since the publication of the extended first law of black hole thermodynamics in \cite{Kastor:2009wy}, which allows for a varying cosmological constant, the addition of pressure and volume terms to obtain a  full cohomogeneity first law has become widespread. In this perspective, the asymptotic spacetime energy $M$ is interpreted as its enthalpy and the cosmological constant acquires the meaning of pressure, with the geometric black hole volume as its conjugate potential. Explicitly, in a $D$-dimensional Schwarzschild-AdS spacetime, the Smarr relation
\begin{equation}
\label{eqn:Smarr1}
(D-3)M= (D-2)TS+2 P V
\end{equation}
and the thermodynamic law
\begin{equation}
\label{eqn:FirstLaw1}
dM= TdS+ VdP
\end{equation}
are obtained using purely geometric methods. These formulas can be extended to rotating and charged cases \cite{Cvetic:2010jb, Dolan:2013ft}, to Taub-NUT spacetimes \cite{Johnson:2014xza, Kubiznak:2019yiu, Ballon:2019uha, Bordo:2019tyh}, to different black hole topologies \cite{Altamirano:2014tva,El-Menoufi:2013pza,Dolan:2010ha} and even to higher curvature theories of gravity \cite{Kastor:2010gq, Liberati:2015xcp}, provided that the notion of thermodynamic volume is appropriately generalized. This quantity will only be equal to the naive spatial black hole volume in the simpler scenarios.  A review of most of these developments can be found in \cite{Kubiznak:2016qmn}.  Additionally, some explicit examples of extended first laws are known to hold in the presence of scalar fields \cite{Cvetic:2010jb,Astefanesei:2019ehu}, in which case the appropriate extension of the volume term is conjectured to be proportional to the average value of the potential inside the black hole region. 

Our study will focus on the thermodynamics of planar black holes  with a minimally coupled scalar field, which are asymptotically anti-de Sitter spacetimes that contain a horizon with the topology of a plane.  It is then wise to first review the case in which the scalar field and any other matter fields are absent. We do this by summarizing some relevant results found in \cite{El-Menoufi:2013pza} and \cite{Traschen:2001pb}. The vacuum planar black hole solution is given by the metric
\begin{equation}
\label{eqn:BBMetric}
ds^2=-\frac{r^2}{\ell^2}\left(1-\frac{\rp^{D-1}}{r^{D-1}}\right)dt^2+\frac{\ell^2}{r^2}\left(1-\frac{\rp^{D-1}}{r^{D-1}}\right)^{-1}dr^2+\frac{r^2}{\ell^2}\delta_{ij}dx^{i}dx^{j}
\end{equation}
where $r=\rp$ is the locus of the horizon and $\ell$ is the AdS radius. Unlike the spherical case, the transverse directions $x_{i}$, $i=1\ldots,D-2$, are not compact, but we can make them so by identifying $x^{i}\equiv x^{i}+L^{i}$. Such a spacetime possesses $(D-1)$ evident Killing vectors $\left\lbrace \partial_{t}, \partial_{i}\right\rbrace$ that generate the time and spatial translations symmetries in each of the $x^{i}$ directions. Associated to each of these symmetries, we can find $(D-1)$ ADM charges relative to the pure AdS background via the prescriptions given, for example in \cite{PhysRevD.31.283}. 

The mass $M$ is the asymptotic charge corresponding to the timelike Killing vector, which reads
\begin{equation}
\label{eqn:MassBB}
M=(D-2)\frac{\rp^{D-1} v}{16 \pi G \ell^2}.
\end{equation}
Here, we have defined the dimensionless quantity $v\equiv \prod_{i=1}^{D-2}\frac{L^{i}}{\ell}$ which is the volume of the box $\mathcal{B}$ with sides parallel to the transversal directions in units of $L$. Similarly, we can calculate the $(D-2)$ ADM tensions corresponding to each of the translation symmetry generators
\begin{equation}
\label{eqn:TensionsBB}
\uptau_{i}=-\frac{\rp^{D-1} v}{16 \pi G \ell^2 L^{i}},
\end{equation}
where these should be interpreted, in fact, as tensions per unit time, given that a time integration is ommitted. We should note that the same results are obtained by using the holographic stress-energy tensor. Thus, the Smarr relation
\begin{equation}
\label{eqn:Smarr2}
M+\sum_{i=1}^{D-2}\uptau_{i}L^i=0
\end{equation}
can be interpreted as the trace of the conformal stress tensor being zero, as is expected for the dual thermal CFT. 

Furthermore, by imposing that the entropy be related to the horizon area via $S=A/4G$ and calculating the temperature with the usual prescription $T=f'(\rp)/4\pi$ we obtain, in terms of $\rp$,
\begin{equation}
\label{eqn:EntropyVac}
S=\frac{\rp^{D-2}v}{4 G}, \qquad T=\frac{D-1}{4}\frac{\rp}{\pi \ell^2}.
\end{equation}
It is then straightforward to verify that, if the AdS scale $\ell$ and the compactifying lengths $L^{i}$ are not regarded as thermodynamic variables, the Smarr relation
\begin{equation}
\label{eqn:Smarr3}
(D-1)M=(D-2)TS
\end{equation}
and the first law of thermodynamics
\begin{equation}
\label{eqn:ProtoFirstLaw}
dM = TdS
\end{equation}
are satisfied. However, in the extended thermodynamics, we do allow a variation in $\ell$ and $L^{i}$ to obtain a differential relation similar to (\ref{eqn:FirstLaw1}). We calculate the geometric volume of the planar black hole by integrating the volume element over the interior
\begin{equation}
\label{eqn:BBVolume}
V=\int_{0}^{\rp}dr\int_{\mathcal{B}}d\vec{x}\sqrt{-g}=\frac{v\rp^{D-1}}{D-1},
\end{equation}
and we identify the spacetime pressure with the cosmological constant 
\begin{equation}
\label{eqn:Pressure}
P=-\frac{\Lambda}{8\pi G} =\frac{(D-1)(D-2)}{16 \pi G \ell^2}.
\end{equation}
Let us observe that
\begin{equation}
\label{eqn:PVM}
PV=M.
\end{equation}
Therefore, we can combine (\ref{eqn:Smarr2}) and (\ref{eqn:Smarr3}) to obtain a Smarr relation that involves all the thermodynamic quantities
\begin{equation}
\label{eqn:Smarr4}
(D-3)M=(D-2)TS + \sum_{i=1}^{D-2}\uptau_{i}L^i - 2 P\Theta
\end{equation}
where the thermodynamic volume $\Theta$ is one half of the geometric volume:
\begin{equation}
\label{eqn:theta}
\Theta\equiv V/2.
\end{equation} 
Note that this Smarr relation, unlike (\ref{eqn:Smarr3}), can be derived from the usual scaling argument according to the engineering dimensions of the thermodynamic charges. With these quantities, the following first law of planar black hole thermodynamics is satisfied  
\begin{equation}
\label{eqn:FirstLaw2}
dM=TdS+\sum_{i=1}^{D-2}\uptau_{i}dL^{i}+\Theta dP. 
\end{equation}
It is interesting to note that, although the extended thermodynamics of black branes has been studied for some time, the relation between the geometric and thermodynamic volumes has never been explicitly stated. This relation is true for planar black holes, but in the spherical case the factor of 1/2 is absent. Although this may seem puzzling, the ultimate reason is that when we compactify the transverse dimensions, the thermodynamic variables depend on the dimensionless transverse volume $v$, which is dependent on $\ell$. Varying the AdS scale will then change the value of $v$. In the spherical case, the transverse coordinates are already compactified so this does contribution does not exist.

The thermodynamics of black branes can be recast without making any reference to the geometric volume; we can express the variable conjugate to the pressure solely in terms of $M$ and $\Lambda$. Namely, using (\ref{eqn:PVM}), we can write
\begin{equation}
\label{eqn:LambdaFirstLaw}
dM=TdS+\sum_{i=1}^{D-2}\uptau_{i}dL^{i}+\frac{M}{2\Lambda}d\Lambda. 
\end{equation}
Armed with this simpler case, we can now proceed to generalize these laws to a spacetime that contains a single scalar field; the generalization to many scalar fields should be straightforward. We will find that the first law as written in (\ref{eqn:LambdaFirstLaw}) applies also to the hairy black branes. An equation analogous to (\ref{eqn:FirstLaw2}) can also be written using an appropriate generalization for the definition of volume. 

In the next section we will review the behaviour of fully back-reacted black brane solutions to the Einstein-Scalar field equations and calculate the asymptotic charges in a very general setting. In section 3 we obtain a Smarr relation akin to (\ref{eqn:Smarr4}) by proposing a generalization for the volume term, which turns out to be similar to that conjectured in \cite{Cvetic:2010jb} and involves a volume integral of the scalar potential between the Poincar\'e and event horizons. We then prove that the proposed quantities satisfy an extended first law. In section 4 we exhibit various examples from the literature and show explicitly that they satisfy the proposed thermodynamic relations. Section 5 is dedicated to the discussion of our results. 

\section{Hairy black brane solutions}

\subsection{Asymptotic behaviour}

We seek to describe the thermodynamics of black branes with scalar hair in a very general setup. This means that we need to characterize some universal asymptotic behaviour for spacetimes in which a scalar field $\phi$  couples minimally to gravity in a $D$-dimensional spacetime. This physical situation is described by the bulk action 
\begin{equation}
\label{eqn:Action}
I_{bulk}=-\int d^{D}x\sqrt{-g}\left(\frac{R}{16\pi G}-\frac{1}{2}\partial_{\mu}\phi\partial^{\mu}\phi-2 \Lambda \mathcal{V}(\phi)\right).
\end{equation}
As can be read from this expression, the scalar potential is taken to be of the form $V(\phi)=-2\Lambda\mathcal{V}(\phi)$ where $\mathcal{V}(\phi)$ is has no explicit dependence on the parameter $\Lambda$. To solve for the metric, we need to find solutions to the Einstein field equations  with a negative cosmological constant in the presence of the scalar field, which gives rise to a non-trivial stress energy tensor.

We are interested in radially symmetric spacetimes, and we will assume that the scalar field respects the same symmetries. With this in mind,  the most general metric ansatz in Poincar\'e coordinates is given by \cite{Gubser:2008ny}
\begin{equation}
\label{eqn:MetricAnsatz}
ds^2=\frac{r^2}{\ell^2}e^{2A(r)}(-h(r)dt^2+\delta_{ij}dx^i dx^j)+\frac{\ell^2}{r^2}\frac{\sigma^2(r)}{h(r)}dr^2, \quad i=2,\ldots, D.
\end{equation}
The constant $\ell$ in this expression is related to the prefactor $\Lambda$ of the potential, which will be varied in the solutions to establish the extended thermodynamics. For asymptotically AdS spacetimes, the relation is
\begin{equation}
\label{eqn:LambdaLength}
\Lambda=-\frac{(D-1)(D-2)}{2 \ell^2};
\end{equation}
however, this relation may change for different asymptotic behaviours of the solutions. An example of this case are the asymptotically dilaton AdS spacetimes \cite{Charmousis:2010zz,Kim:2012pd,Kastor:2018cqc}, where the potential is given by an exponential function of a scalar field with a finite value at infinity. In this work, we will restrict our study to asymptotically AdS spacetimes. In terms of the potential, this means that $\mathcal{V}(\phi_{\infty})=1/16\pi G$, where $\phi_{\infty}$ is the limit of the value of the scalar field as one approaches the conformal boundary.

In terms of the functions $A(r)$, $\sigma(r)$ and $h(r)$, there are three independent gravitational field equations which read
\begin{equation}
\label{eqn:eom1}
(D-2) \left(r A'+1\right) \left((D-1) h \left(r A'+1\right)+r h'\right)-8\pi G(r^2 h \phi'^2-2 \ell^2 V(\phi)\sigma^2)=0,
\end{equation}
\begin{equation}
\label{eqn:eom2}
\left(((D-1) r A'+D) -\frac{r\sigma'}{\sigma}\right) h'+r h''=0,
\end{equation}
\begin{equation}
\label{eqn:eom3}
r \left((D-2) A''+8 \pi G \phi'^2\right)+(D-2) A'-(D-2)(1+rA')\frac{\sigma'}{\sigma}=0.
\end{equation}
The condition that the spacetime be asymptotically AdS translates into the limits $A\rightarrow 0 $, $\sigma\rightarrow 1 $ and $h\rightarrow 1$ as $r\rightarrow \infty$. In general, it is no trivial matter to obtain solutions to these coupled equations. The handful of solutions that have been found in the literature use a specific form of the potential that allow for exact solutions (see, for example: \cite{Acena:2012mr, Acena:2013jya,Fan:2015ykb}). Nevertheless, we can make some progress as equation (\ref{eqn:eom2}) allows us to express the blackening factor $h$  as the exact integral, independent of the form of the potential. We obtain
\begin{equation}
\label{eqn:BlackFactor}
h(r)=(D-1)\mu\int_{r_{+}}^{r}\frac{e^{-(D-1)A(r)}\sigma(r)}{r^D}dr,
\end{equation}
where the constant $\mu$ is chosen so that $h\rightarrow 1$ as $r\rightarrow \infty$. The radial value $r_{+}$ is a zero of $h$, which is required to exist if we are to have a horizon at $r=r_{+}$. Note that, at large values of  $r$, we can expand 
\begin{equation}
\label{eqn:AsympBlackFactor}
h\sim 1-\frac{\mu}{r^{D-1}}+\ldots
\end{equation}

The full characterization of the asymptotics of the metric will depend on the particular form of the scalar potential. Without loss of generality, if $\phi\rightarrow 0$ as $r\rightarrow\infty$, the asymptotic behaviour of the scalar field to leading order in $r$ is
\begin{equation}
\label{eqn:AsympPhi}
\phi\sim \frac{\phi_{0}}{r^{D-1-\Delta}}+\frac{\phi_{+}}{r^{\Delta}}.
\end{equation}
Here, the exponent $\Delta$ is fixed by the specific form of the potential and corresponds holographically to the conformal dimension of the dual operator. In the following, this asymptotic behaviour is taken to be fixed.

In the above, we have not made any assumptions about the behaviour of the metric close to $r=0$. In section 3, we will impose a certain condition on the behaviour of the functions so that there is a singularity at $r=0$ and the integral of the potential is finite inside the horizon. 

\subsection{ADM charges}
 
We face a challenge when calculating the masses of scalar field spacetimes using the ADM method: if we have an asymptotic behaviour of the scalar field characterized by two  or more parameters, then it has been shown in \cite{Lu:2013ura,Lu:2014maa} that the Hamiltonian variation at infinity cannot be expressed as a total derivative, so the mass is not well defined. We will circumvent this problem by taking the pure scalar field spacetime as the background, and calculate the mass contribution of the planar black hole only. Fot this purpose, we will use a version of the Hamiltonian formalism introduced in \cite{PhysRevD.31.283}. 

To calculate the ADM mass, first we foliate spacetime with spacelike hypersurfaces $\Sigma$, with unit normal vector $n^{\mu}$. The induced metric in these hypersurface is then 
\begin{equation}
\label{eqn:InducedMetric}
\gamma_{\mu\nu}=g_{\mu\nu}-n_{\mu}n_{\nu}.
\end{equation}
The canonical momentum associated to $\gamma_{\mu\nu}$ is 
\begin{equation}
\label{eqn:ConjugateMomentum}
\pi^{\mu\nu}=-\sqrt{\gamma}\left(K\gamma^{\mu\nu}-K^{\mu\nu}\right)
\end{equation}
where $K_{\mu\nu}$ is the extrinsic curvature of $\Sigma$. The second ingredient in the calculation is a background metric, which we take to be the pure scalar field spacetime, defined by setting $h(r)=1$ in (\ref{eqn:MetricAnsatz}). The induced background metric in the boundary and its conjugate momentum are denoted by $\gamma^{(0)}_{\mu\nu}$ and $\pi^{(0)}_{\mu\nu}$ respectively. The corresponding differences between the quantities on $\Sigma$ and on the background are denoted by $\delta \gamma_{\mu\nu}$ and $\delta \pi_{\mu\nu}$. 

The ADM mass is the conserved charge associated to the conserved Noether current correspondent to the time translation symmetry of our spacetime. The background Killing vector that generates this symmetry is $\xi=\partial/\partial t$. If $F$ is the lapse of $\xi$, meaning the component of $\xi$ along the direction of the normal vector field to the foliation $\Sigma$, and $F^{a}$ is the shift, the component parallel to the family of hypersurfaces, then the mass can be expressed as 
\begin{equation}
\label{eqn:ADMCharge}
M=-\frac{1}{16 \pi G}\int_{\partial\Sigma_{\infty}}da_{c}B^{c}
\end{equation}
where \cite{PhysRevD.31.283}
\begin{equation}
\label{eqn:Bvector}
B^{c}= F(D^{c}\delta\gamma-D_{b}\delta\gamma^{bc})-\delta\gamma D^{c}F+\delta\gamma^{cb}D_{b}F +\frac{1}{|\gamma|}F^{d}\left( \pi^{(0)ab}\delta\gamma_{ab}\tensor{\gamma}{^{(0)}^c_d}+2\pi^{(0)cb}\delta\gamma_{db}-2\delta\tensor{\pi}{^c_d} \right).
\end{equation}
Here, $D_{a}$ denotes the covariant derivative corresponding to the background induced metric $s^{(0)}_{ab}$. We note that, in general, there is an additional contribution of matter to the vector $B^{c}$. For our purposes, we will hold the asymptotic behaviour of the scalar field fixed, so that $\delta\phi=0$. We can take a foliation $\Sigma$ such that the normal vector field to the surfaces is parallel to the timelike Killing vector $\xi=\partial/\partial t$. As a consequence, $\xi$ has lapse $F=r/L$ and shift $F^{a}=0$. In our case, the perturbations read $\delta\gamma_{tt}=\delta\gamma_{rr}=\mu/r^{(D-1)}$. Evaluating the mass of the black brane we obtain, in $D$ spacetime dimensions
\begin{equation}
\label{eqn:ADMMass}
M=(D-2)\frac{\mu v}{16 \pi G \ell^2}.
\end{equation}

The tensions can be calculated in a similar way \cite{Traschen:2001pb}. The only difference is that the foliation of spacetime is now given by timelike hypersurfaces (changing the minus sign in (\ref{eqn:InducedMetric})). This foliation is chosen in such a way that the normal vector is parallel to the Killing vector $\xi_{k}=\partial/\partial x^{k}$. For the spatial tension associated to the symmetry in the $k$-th spatial direction we obtain
\begin{equation}
\label{eqn:ADMTension}
\uptau_{k}=-\frac{\mu v}{16 \pi G \ell^2 L^{k}},
\end{equation}
As in the case of the hairless black brane, we find that the mass and the tensions satisfy
\begin{equation}
\label{eqn:Traceless}
M+\sum_{i=1}^{D-2}\uptau_{k}L^{k}=0.
\end{equation}
Having assumed that the spacetime is asymptotically AdS, this is again interpreted as the trace of the stress energy tensor in the dual CFT being zero\footnote{The dual to a scalar black brane spacetime is a QFT that comes from a deformed CFT. This traceless property refers to the stress energy tensor of the non-deformed theory.}. We do not expect this to hold in the case where the asymptotics deviate from anti-de Sitter. Such hyperscaling violating spacetimes and their thermodynamics have been studied, for example, in the case of asymptotically dilaton AdS spacetimes \cite{Kastor:2018cqc}. and for a hyperscaling violating generalization of Lifshitz black holes \cite{Pedraza:2018eey}. 

It is worthwhile to insist on the fact that the mass calculated above does not include the energy contribution from the scalar field; as such it represents a gravitational energy. To account for the energy contained in the scalar field, one would have to use either the holographic \cite{Brown:1992br,Balasubramanian:1999re} or Wald prescriptions \cite{Wald:1993nt} using appropriate counterterms \cite{Emparan:1999pm, Buchel:2013lla, Papadimitriou:2005ii}. In this paper we do not vary the asymptotic behaviour of the scalar field, so it will not contribute to the first law of thermodynamics. Work in these lines, albeit without varying the cosmological constant, has been carried out in \cite{Liu:2015tqa, Gursoy:2018umf}. 

\section{Extended thermodynamics}

\subsection{Smarr relation}
 We now proceed to  derive a Smarr relation for the scalar-black brane solution. Such an equation will relate the mass to the entropy, the tensions and the black brane volume, which is in general different from the vacuum black brane spacetime. Obtaining such an equation tells us what to expect the thermodynamic volume to be equal to, so that we can check the first law of thermodynamics  later. There are two ways to go about this: the first is to perform a scaling transformation and use Euler's theorem, which is straightforward provided that we already know the first law; the second is to use geometric methods, namely a Komar intergration. In this section we will illustrate how to derive the Smarr relation using the scaling argument. 
  
The scaling method was already presented in \cite{Liu:2015tqa} in the context of a fixed cosmological constant. We utilize this result in order to generalize it to varying compactifying and AdS scales. Under a scaling transformation. Without rescaling $\ell$, there exists a scaling symmetry of the metric
\begin{equation}
\label{eqn:Scaling1}
r\rightarrow \lambda r, \qquad t\rightarrow \lambda^{-1}t, \qquad x^{i}\rightarrow x^{i}\lambda^{-1}
\end{equation}
provided that $\mu\rightarrow \lambda^{D-1}\mu$. The entropy scales as $S\rightarrow \lambda^{D-2}S$, as can be deduced from dimensional analysis. Considering the mass as a function only of the entropy, we observe that $M(S)$ is a homogeneous function such that
\begin{equation}
\label{eqn:MassScaling}
M(\lambda^{D-2}S)=\lambda^{D-1}M(S).
\end{equation}

Euler's theorem for homogeneous functions states that, if a function $f$ is such that 
\begin{equation}
f(\lambda^{p}x, \lambda^{q}y,\lambda^{r}z)=\lambda^{s}f(x,y,z),
\end{equation}
 then
\begin{equation}
\label{eqn:EulerTheorem}
s f = p\left(\frac{\partial f}{\partial x}\right)+q\left(\frac{\partial f}{\partial y}\right)+r\left(\frac{\partial f}{\partial z}\right).
\end{equation}
The homogeneity of the mass along with this theorem imply that we must have
\begin{equation}
\label{eqn:ProtoSmarr}
(D-1)M=(D-2)\left(\frac{\partial M}{\partial S}\right)S.
\end{equation}
This relation still holds true once we vary the compactifying and AdS scales. However, it is not useful to define a thermodynamic volume. First of all, one would need to know a priori that $\partial M/\partial S$ is the temperature to obtain a Smarr relation as in equation (\ref{eqn:Smarr3}), and for that we need to know that a first law $dM=TdS$ holds true. This is indeed the case, as shown in \cite{Liu:2015tqa}. Secondly, we need to study the mass dependence on the cosmological constant to obtain a relation akin to (\ref{eqn:Smarr4}). We can do this by noticing that the scaling transformation
\begin{equation}
\label{eqn:Scaling2}
r\rightarrow \lambda r, \qquad t\rightarrow \lambda t, \qquad x^{i}\rightarrow \lambda x^{i}, \qquad \ell \rightarrow \lambda \ell
\end{equation}
is a symmetry of the metric. Note that because of the transformation rule for the transversal coordinates $x^{i}$, the compactifying lengths scale as $L^{k}\rightarrow \lambda L^{k}$. 

Considering the mass as a function of the $L^{k}$ and the cosmological constant, we see that it is homogeneous in the following way:
\begin{equation}
\label{eqn:MassScaling2}
M(\lambda^{D-2}S, \lambda L^{k}, \lambda^{-2}\Lambda)=\lambda^{D-3}M(S,L_{k},\Lambda).
\end{equation}
Then Euler's theorem immediately implies that
\begin{equation}
\label{eqn:EulerSmarr1}
(D-3)M=(D-2)\left(\frac{\partial M}{\partial S}\right)S+\sum_{k}\left(\frac{\partial M}{\partial L^{k}}\right)L^{k}-2\left(\frac{\partial{M}}{\partial\Lambda}\right)\Lambda.
\end{equation}
Now, we recall that $M+\sum_{T_{k}}L_{k}=0$ from the fact that the holographic stress energy tensor must be traceless. Then, by subtracting $2M$ on each side of equation (\ref{eqn:ProtoSmarr}), it is permissible to write
\begin{equation}
\label{eqn:FinalSmarr}
(D-3)M=(D-2)\left(\frac{\partial M}{\partial S}\right)S+\sum_{k}\uptau_{k}L^{k}-2\left(\frac{M}{2\Lambda}\right)\Lambda.
\end{equation}
From this we could identify, up to some overall coefficients, $\partial M/\partial L^{k}=\uptau^{k}$ and $\partial M/\partial \Lambda =M/2\Lambda$, the latter being proportional to the volume term. This turns out to be the correct identification but an independent derivation of the first law is needed to make it precise.

\subsection{Komar Integration}

Before deriving such a first law, we present a geometric derivation of the Smarr relation using Komar integration. This derivation will elucidate a relation between the integral of the potential behind the black brane horizon and the volume term. By tracing out the gravitational field equations, we can express the Ricci tensor as
\begin{equation}
\label{eqn:Ricci}
R_{\mu\nu}=8 \pi G \left(\partial_{\mu}\phi\partial_{\nu}\phi + \frac{2}{D-2}g_{\mu\nu}V(\phi)\right);
\end{equation}
then, we contract this equation with the timelike Killing vector $\xi^{\alpha}$ and, using the Killing identity $\square\xi^{\alpha}=-\tensor{R}{^\alpha_\beta}\xi^{\beta}$, we observe that
\begin{equation}
\label{eqn:GaussLaw}
\nabla_{\mu}\nabla^{\mu}\xi^{\nu}+\frac{16 \pi G}{D-2}\xi^{\nu}V(\phi)=0.
\end{equation}

We have assumed that the scalar field is static, so that $\xi^{\alpha}\partial_{\alpha}\phi=0$. This also implies that the divergence $\nabla_{\alpha}(\xi^{\alpha}V(\phi))$ vanishes. Therefore, there exists a Killing potential, which is a 2-form $\omega^{\alpha\beta}$ such that
\begin{equation}
\label{eqn:KillingPotential}
V(\phi)\xi^{\beta}=\nabla_{\alpha}(V(\phi)\omega^{\alpha\beta}). 
\end{equation}
Combining the previous results, we have that 
\begin{equation}
\label{eqn:GaussLaw2}
\nabla_{\mu}\left(\nabla^{\mu}\xi^{\nu}+\frac{16 \pi G}{D-2}\omega^{\mu\nu}V(\phi)\right)=0.
\end{equation}
We can then calculate the integral of this vanishing 1-form over a codimension-1 spacelike hypersurface $\Sigma$ and, by virtue of Stokes' theorem, we may express this result as
\begin{equation}
\label{eqn:KomarIntegral}
I=\int_{\partial\Sigma}d\sigma_{\mu\nu}\left(\nabla^{\mu}\xi^{\nu}+\frac{16 \pi G}{D-2}\omega^{\mu\nu}V(\phi)\right)=0.
\end{equation}
where $\sigma_{\mu\nu}$ is the binormal to the boundary of $\Sigma$, given a choice of orientation. The boundary $\partial\Sigma$ has a component at infinity and a component at the bifurcation surface $H$. Hence, we can cast this as a sum $I_{\infty}-I_{H}=0$. 

The assumption of spherical symmetry implies that the only non-vanishing components of $\omega^{\alpha\beta}$ are $\omega^{rt}=-\omega^{tr}$. In fact, we can use equation  (\ref{eqn:KillingPotential}) to express this component in terms of an integral of the potential
\begin{equation}
\label{eqn:KillingPotentialSol}
\omega^{rt}=\frac{1}{\sqrt{-g}V(\phi)}\int_{\rho}^{r} dr' \sqrt{-g}V(\phi).
\end{equation}
Here $\rho$ is a constant of integration and is therefore arbitrary, and it accounts for the fact that the Killing potential is unique only up to the addition of a closed 2-form. The integral (\ref{eqn:KomarIntegral}) can then be rewritten as
\begin{equation}
\label{eqn:KomarIntegral2}
I=\int d^{D-2}x\sqrt{-g}\nabla^{r}\xi{^t}+\frac{16 \pi G}{D-2}\int d^{D-2}x\int_{\rho}^{r}dr'\sqrt{-g} V(\phi)=0.
\end{equation}
We need to evaluate this integral at infinity and at the horizon. Both integrals will depend on our choice of $\rho$, but we expect this dependence to cancel out. Remarkably, there is an exact expression for the integral $I$ that comes from using the equations of motion. First, one must use (\ref{eqn:eom3}) to find $\phi'$ in terms of the metric functions $A$ and $\sigma$. This result can be plugged into equation (\ref{eqn:eom1}) to solve for the potential. Then the quantity $\sqrt{-g}V(\phi)$ is a total derivative in the radial variable, so the indefinite integral can be readily calculated. As a result we obtain
\begin{equation}
\label{eqn:W}
W(r)\equiv\int dr\sqrt{-g}V(\phi)= -\frac{(D-2) r^{D-1} h(r) e^{(D-1) A(r)} \left(r A'(r)+1\right)}{16 \pi  G \ell^D \sigma(r)}.
\end{equation}
There are no integrals over the radial coordinate for the $\nabla^{r}\xi^{t}$ term, so this term can be evaluated without any further complications. The result is
\begin{equation}
\nabla^{r}\xi^{t}= \frac{r \left(2 h(r) \left(r A'(r)+1\right)+r h'(r)\right)}{2\ell^2 \sigma^2 (r)}.
\end{equation}

We can use these results to get a simple expression for the integrand in (\ref{eqn:KomarIntegral2}). We obtain
\begin{equation}
\label{eqn:IntegralR}
I=\frac{r^D e^{(D-1)A(r)} h'(r)}{2 \ell^2 \sigma(r)} v-\frac{16 \pi G W(\rho)}{D-2}v
\end{equation}
With this expression at our disposal, we can calculate $I_{\infty}$ at a sphere at infinity and $I_{H}$ at the horizon. Here, we note that $W(r_{+})=0$ provided that $\sigma(r)$, $A(r)$ and its derivative are finite at the horizon, that is, we assume that the coordinate singularity at $r=r_{+}$ comes only from the blackening factor $h(r)$. We also use the fact that $\xi^{\alpha}$ generates the black brane horizon, and hence the integral of $\nabla^{\mu}\xi^{\nu}$ over $H$ is equal to $8\pi G TS$.  The asymptotic behaviours of $A$ and $h$ allow us to express
\begin{equation}
\label{eqn:IntegralInfty}
I_{\infty}= 8 \pi G \left( \frac{D-1}{D-2}M-\frac{2 W(\rho) v}{D-2}\right),
\end{equation}
\begin{equation}
\label{eqn:IntegralHorizon}
I_{H}= 8 \pi G \left( T S -\frac{2 W(\rho) v}{D-2}\right).
\end{equation}
Upon imposing $I_{\infty}=I_{H}$, the $\rho$ dependent terms cancel. We also note that the divergent part of the integral of $\nabla^{r}\xi^{t}$ is always cancelled by the integral of the potential, yielding a term proportional to the mass. We thus obtain the Smarr relation
\begin{equation}
\label{eqn:FakeSmarr}
(D-1)M=(D-2)TS,
\end{equation}
which is the same as (\ref{eqn:Smarr3}). Thus, considering $\Lambda$ and $L_{k}$ thermodynamic variables as before and defining $P=-\Lambda/8 \pi G$, we obtain the following Smarr relation analogous to (\ref{eqn:Smarr4}):
\begin{equation}
\label{eqn:FinalSmarr2}
(D-3)M=(D-2)TS + \sum_{i=1}^{D-2}\uptau_{i}L^i - 2 P\Theta.
\end{equation} 
Here, we have subtracted $2M$ on each side of (\ref{eqn:FakeSmarr}) by using (\ref{eqn:Traceless}) and by defining the thermodynamic volume
\begin{equation}
 \Theta\equiv -4\pi G M/\Lambda. 
\end{equation} 
This means that the extended thermodynamics of black branes is in fact very simple, and the thermodynamic volume can be readily calculated once we know the mass.  Comparing to the purely gravitational case, this makes us wonder what is the relation  between this thermodynamic volume and the actual geometric volume enclosed between the singularity and the black brane horizon. This relation was straightforward in the hairless case: all one had to do is to calculate both volumes in terms of $r_{+}$ and compare, then we observed that the thermodynamic volume was one half of the geometric one. In our present case, this relation is a bit more complicated but, as we will see, the integral of the potential over the volume behind the horizon turns out to be proportional to the mass, from which $\Theta$ can be readily calculated. The constant of proportionality depends on the particular form of the solution but has a simple expression when the metric is expressed in Poincar\'e coordinates. Consider the integral
\begin{equation}
\label{eqn:Upsilon}
\Upsilon = \int_{\mathcal{B}}d^{D-2}x \int_{0}^{r_{+}}dr V(\phi)=(W(r_{+})-W(0))v.
\end{equation}
We know that $W(r_{+})$ is zero, so all that remains to do is to calculate $W(0)$. This may seem complicated because we did not impose a particular behaviour of the metric in the deep infrarred. However, under certain conditions, we can simplify its calculation. First, we need to assume that both $A(r)$ and its derivative are finite as $r$ goes to zero. Secondly, we assume that $\sigma(r)\sim 1/r^{\alpha}$, $\alpha \geq 1-D$ as $r\rightarrow 0$. This last assumption ensures that the integral $\Upsilon$ is finite and that there is a singularity of the metric at $r=0$. We can then calculate the limit using the exact formula for $h(r)$:

\begin{align}
W(0)&=\lim_{r\rightarrow 0} \left(-\frac{(D-2) r^{D-1} h(r) e^{(D-1) A(r)} \left(r A'(r)+1\right)}{16 \pi  G \ell^D \sigma(r)}\right)\nonumber \\
 &=-\left(\frac{(D-2) e^{(D-1) A(0)}}{16 \pi  G \ell^D}\right)\lim_{r\rightarrow 0}\frac{h(r)}{\sigma(r)/r^{D-1}}. \label{eqn:Wzero1}
\end{align}
We observe that when $\alpha\geq 1-D$, then both $h$ and $\sigma/r^{D-1}$ diverge as $r$ approaches zero. This means that we are allowed to use L'Hopital's rule to evaluate the above limit.

\begin{align}
W(0)&=-\left(\frac{(D-2) e^{(D-1) A(0)}}{16 \pi  G \ell^D}\right)\lim_{r\rightarrow 0}h'(r)\left(\frac{\sigma'(r)}{r^{D-1}}-(D-1)\frac{\sigma(r)}{r^D}\right)^{-1}\nonumber \\
&=-\left(\frac{(D-2)(D-1)\mu  e^{(D-1) A(0)}}{16 \pi  G \ell^D}\right)\lim_{r\rightarrow 0}\frac{e^{-(D-1)A(r)}\sigma(r)}{r^{D}}\left(\frac{\sigma'(r)}{r^{D-1}}-(D-1)\frac{\sigma(r)}{r^D}\right)^{-1}\nonumber \\
&=\frac{M}{v}\lim_{r\rightarrow 0}\left(1-\frac{1}{D-1}\frac{r\sigma'(r)}{\sigma(r)}\right)^{-1}=\frac{M}{v}\left(\frac{D-1}{D-1+\alpha}\right).\label{eqn:Wzero2}
\end{align}
Therefore, the integral $\Upsilon$ becomes
\begin{equation}
\label{eqn:UpsilonM}
\Upsilon=-\left(\frac{D-1}{D-1+\alpha}\right)M.
\end{equation}
Plugging this into the thermodynamic volume $\Theta=-4\pi G M/\Lambda$, we conclude that
\begin{equation}
\label{eqn:ThetaPotential}
\Theta=8 \pi G \left(1+\frac{\alpha}{D-1}\right)\int_{\mathcal{B}}d^{D-2}x\int_{0}^{r_{+}}dr\mathcal{V}(\phi).
\end{equation}
This provides evidence for the conjecture that the thermodynamic volume must be proportional to the integral of the potential behind the horizon in the case of a hairy black hole.  However, we are not finished yet. We need to prove that we can obtain an extended first law using this proposed definition of thermodynamic volume. This is the purpose of the following subsection.

\subsection{Extended first law}

The Hamiltonian method that we previously employed to calculate the mass and the tensions can also be used to derive the first law of thermodynamics. The trick is to take the scalar black brane spacetime as the background spacetime, and then calculate the mass of a the spacetime given by a metric with the parameters $\mu$, $\ell$ and $L_{k}$ slightly shifted. This will give us a relation between the perturbed parameters  $\delta\mu$, $\delta\ell $ and $\delta L_{k}$, and from here we can obtain the first law of thermodynamics. This procedure is known as the  Hamiltonian perturbations formalism \cite{ PhysRevD.31.283, Kastor:2009wy, Kastor:2018cqc}.

When we have a variable cosmological constant, this procedure needs to be done with more care, as  now the Hamiltonian for the scalar fields needs to be taken into account. This is due to the fact that, even though we keep the asymptotic behaviour of the scalar field constant, the potential has an explicit overall factor of $\Lambda$ that needs to be accounted for in the perturbations. The linearized perturbations can be shown to satisfy the Einstein field equations; as a consequence, on a constant time slice $\Sigma$,
\begin{equation}
\label{eqn:GaussTherm}
D_{a}\left(\frac{B^{a}}{16 \pi G}+2\delta\Lambda\mathcal{V}(\phi)\omega^{ab}n_{b}\right)=0.
\end{equation}
Here $B^{a}$ is calculated as in (\ref{eqn:Bvector}) and we use the same notation as in section II B. This time, the metric perturbations  as $r$ goes to infinity are given by
\begin{equation}
\label{eqn:PertTherm1}
\delta\gamma_{rr}=\left(\frac{\ell}{r}\right)^2\frac{\delta\mu}{r^{D-1}},
\end{equation}
\begin{equation}
\label{eqn:PertTherm2}
\delta\gamma_{ij}=\delta_{ij}\frac{r^2}{\ell^2}\left(2\frac{\delta L_{k}}{L_{k}}-2\frac{\delta\ell}{\ell}\right).
\end{equation}
To establish the first law, we integrate (\ref{eqn:GaussTherm}) over $\Sigma$ and use Stokes' theorem to express it as a boundary integral. As before, the boundary of $\Sigma$ has disconnected components $\partial\Sigma_{\infty}$ at infinity and $\partial\Sigma_{H}$ at the horizon. We define
\begin{equation}
\label{eqn:JInfinity}
J_{\infty}=\int_{\partial\Sigma_{\infty}}d\sigma_{a}\left(\frac{B^{a}}{16 \pi G}+2\delta\Lambda\mathcal{V}(\phi)\omega^{ab}n_{b}\right), \quad \textrm{and}
\end{equation}
\begin{equation}
\label{eqn:JHorizon}
J_{H}=\int_{\partial\Sigma_H}d\sigma_{a}\left(\frac{B^{a}}{16 \pi G}+2\delta\Lambda\mathcal{V}(\phi)\omega^{ab}n_{b}\right);
\end{equation}
therefore we have $J_{\infty}-J_{H}=0$. For our purposes, it is easier to express the potential part of the integral as a volume integral. This can be done by virtue of (\ref{eqn:KillingPotentialSol}), so that

\begin{equation}
\label{eqn:GaussTherm2}
\int_{\partial\Sigma_{\infty}}d\sigma_{a}\left(\frac{B^{a}}{16 \pi G}\right)+2\delta\Lambda\int d^{D-2}x\int_{r_{+}}^{\infty}dr\sqrt{-g}\mathcal{V}(\phi)=\int_{\partial\Sigma_{H}}d\sigma_{a}\left(\frac{B^{a}}{16 \pi G}\right).
\end{equation}

This allows us to use the exact expression for the integral of the potential that we calculated previously in (\ref{eqn:W}). It only remains to calculate all these integrals and express the variations in terms of $\delta M$. As shown in \cite{PhysRevD.46.1453}, the integral at the horizon evaluates to $-T\delta S$, no matter what gauge fields are present Lagrangian. We proceed to evaluate, for a fixed large $r$ hypersurface
\begin{equation}
\label{eqn:HPert1}
\int_{\partial\Sigma_{r}}d\sigma_{a}\left(\frac{B^{a}}{16 \pi G}\right)= -\frac{(D-2)}{8 \pi G} \frac{r^{D-1}}{\ell^{D+1}}\delta\ell+\frac{M}{\ell}\delta \ell-\delta M-\sum_{k=1}^{D-2}\uptau_{k}\delta L^{k},
\end{equation}
\begin{equation}
\label{eqn:HPert2}
2\delta\Lambda\int d^{D-2}x\int_{r_{+}}^{\infty}dr\sqrt{-g}\mathcal{V}(\phi)=\frac{(D-2)}{8 \pi G} \frac{r^{D-1}}{\ell^{D+1}}\delta\ell-\frac{2M}{\ell}\delta\ell.
\end{equation}
Putting all of this together and expressing the variations of $\ell$ in terms of $\delta\Lambda$, we get the first law for scalar black brane spacetimes
\begin{equation}
\label{eqn:FinalFirstLaw}
dM=TdS+\sum_{k=1}^{D-2}\uptau_{k}L^{k}+\frac{M}{2\Lambda}d\Lambda,
\end{equation}
which is the same as in the vacuum solution (\ref{eqn:BBMetric}). Defining the pressure $P$ as in (\ref{eqn:Pressure}) and the thermodynamic volume $\Theta=-4\pi G M/\Lambda$ , we obtain
\begin{equation}
\label{eqn:FinalFirstLaw2}
dM=TdS+\sum_{k=1}^{D-2}\uptau_{k}L^{k}+\Theta dP.
\end{equation}

As we see, the only difference between the hairy and non-hairy cases is that the thermodynamic volume $\Theta$ is related in a different way to the natural generalization of geometric volume
\begin{equation}
\label{eqn:GeometricVolume}
V=16\pi G \int_{\mathcal{B}}d^{D-2}x\int_{0}^{r_{+}}dr\mathcal{V}(\phi).
\end{equation}
From equation (\ref{eqn:ThetaPotential}) we have, more precisely, that the relation between these two volumes is
\begin{equation}
\label{eqn:GeometricThermVolume}
\Theta=\frac{1}{2}\left(1+\frac{\alpha}{D-1}\right)V,
\end{equation}
where we recall that $\sigma\sim 1/r^{\alpha}$ close to $r=0$, so it characterizes the infrared behaviour of the metric. This geometric volume\footnote{The hairless case can be considered to be a solution with a constant scalar field and a potential equal to $2\Lambda/16\pi G$. This is the reason why the prefactor of $16\pi G$ is needed in this definition. Let us observe that when $\alpha=0$, the thermodynamic volume is half of the geometric volume, as expected.} is in general difficult to calculate as, in the explicit solutions, the expressions for the scalar potential are quite complicated. It is remarkable that there is such a simple expression for this integral of the potential in terms of the mass, but we must note that such a relation is not that straightforward in the case of spherical black holes. We leave this analysis for future work.
\section{Application to explicit solutions}

\subsection{Hypergeometric solution}

Having derived a first law and a Smarr relation for hairy black branes, an obvious sanity check is to verify that these hold for some explicit solutions. The Einstein field equations are not easy to solve exactly and moreover, it is only possible to do so when the form of the potential allows for an explicit  integration of the metric. When this is possible, the constants of integration give us a family of solutions. In our sample metrics, we will also verify the relation (\ref{eqn:GeometricThermVolume}) between the thermodynamic and geometric volumes in some particular cases where the integration of the potential is not too complicated.

The first sample solution that we will study was derived in \cite{Fan:2015ykb}, which we review here. The potential is given by\footnote{This potential and the following one are slightly adapted from the original source so that they are of the form $V(\phi)=2\Lambda \mathcal{V}(\phi)$.}
\begin{align}
16 \pi G V(\phi)= & 2\Lambda(\cosh\phi)^{\frac{\nu k_{0}^2}{D-2}}\left(1-\eta(\sinh\phi)^{\frac{D-1}{\nu}}{}_{2}F_{1}\left(\frac{D-1}{2\nu},\frac{\nu k_{0}^2}{4(D-2)},\frac{D+2\nu+1}{2\nu},\sinh^2\phi\right)\right)\nonumber \\
& \times\left(1-\frac{\nu^2 k_{0}^2 \tanh^2\phi}{2(D-1)(D-2)}\right)+2\Lambda\eta(\cosh\phi)^{\frac{\nu k_{0}^2}{2(D-2)}}(\sinh\phi)^{\frac{D-1}{\nu}}.\label{eqn:PotentialHG}
\end{align}
Here $\eta$, $k_{0}$ and $\nu$ are numerical constants that parametrize this class of solutions. The constant $\nu$ is required to be in the interval $1/2(D-1)<\nu<(D-1)$ so that the mass of the scalar  satisfies the BF stability condition and the potential reaches a local maximum at $\phi=0$. With this potential, it is straightforward to verify that, by plugging in $A(r)=0$ and the following functions in our ansatz (\ref{eqn:MetricAnsatz})
\begin{equation}
\sigma(r)=\left(1+\frac{q^{2\nu}}{r^{2\nu}}\right)^{-\frac{\nu k_{0}^2}{4(D-2)}}, \quad h(r)=1-\frac{q^{D-1}\eta}{r^{D-1}}{}_{2}F_{1}\left(\frac{D-1}{2\nu},\frac{\nu k_{0}^2}{4(D-2)},\frac{D+2\nu+1}{2\nu},-\frac{q^{2\nu}}{r^{2\nu}}\right) \label{eqn:MetricHG}
\end{equation}
and with the scalar field 
\begin{equation}
\label{eqn:ScalarHG}
\phi=\sinh^{-1}\left(\frac{q^{\nu}}{r^{\nu}}\right)
\end{equation}
then the Einstein field equations are satisfied. By expanding the function $h(r)$ near infinity, the mass and the tensions and the thermodynamic volume can be read off:
\begin{equation}
\label{eqn:ChargesHG}
M=(D-2)\frac{\eta q^{D-1}}{16 \pi G \ell^2}v, \quad \uptau_{k}=-\frac{ \eta q^{D-1}}{16 \pi G L_{k}\ell^2}v, \quad \Theta= \frac{\eta q^{D-1}}{2(D-1)}v
\end{equation}
The temperature $T$ and the entropy $S$ can be found using the usual methods. We find
\begin{equation}
\label{eqn:TempEntropHG}
T=\frac{(D-1)\eta q^{D-1}}{4\pi r_{+}^{D-2}\ell^2}, \quad S=\frac{r_{+}^{D-2}v}{4},
\end{equation}
so indeed, we verify that $(D-1)M=(D-2)TS$ and, with $P=-\Lambda/(8\pi G)$, we also obtain the extended Smarr relation (\ref{eqn:FinalSmarr2}). To derive the first law, we observe that $h$ is a function of $q/r$ so, given that $h(r_{+})=0$, then $q/r_{+}$ must be a constant and the variations are related via 
\begin{equation}
\label{eqn:SmartIdeaHG}
\delta r_{+}= \frac{r_{+}}{q}\delta q. 
\end{equation}
Using this fact, the extended first law (\ref{eqn:FinalFirstLaw2}) can be readily verified. 

Now we would like to find the relation between the geometric and thermodynamic volumes. As we can see, the volume integral of the potential is not trivial to carry out. As an example, we use $D=5$, $\nu=2$ and $k_{0}=1$, given that the integral simplifies significantly in this case\footnote{i.e. Mathematica can do it.}. Expressing the potential in terms of the coordinate $r$, we have
\begin{equation}
\label{eqn:PotentialHG2}
16\pi G V(\phi(r))=-\frac{\Lambda  \left(36 \eta  r^{2/3} \left(q^4+r^4\right)^{5/6}-(6 \eta +5) \left(5 q^4+6 r^4\right)\right)}{30 r^{4/3} \left(q^4+r^4\right)^{2/3}}.
\end{equation}
From this expression, we can calculate the indefinite volume integral of the potential
\begin{equation}
\label{eqn:IntegralPotentialHG}
16\pi G \int d^{3}x\int dr \sqrt{-g}V(\phi(r))=-\frac{2}{5\ell^2}\left(\frac{3}{2} (6 \eta +5) r^{10/3} \sqrt[6]{q^4+r^4}-9 \eta  r^4\right)v.
\end{equation}
We would like to integrate this from $r=0$ to $r=r_{+}$. The antiderivative evaluates to zero at $r=0$; to evaluate at $r=r_{+}$, we can find the value of $q^4$ from the equation $h(r_{+})=0$. The hypergeometric function simplifies in this case and we get
\begin{equation}
\label{eqn:q4HG}
q^4=\frac{\left(5\times 6^{4/5} \sqrt[5]{\frac{5}{\eta }+6}+\left( 6^{9/5} \sqrt[5]{\frac{5}{\eta }+6}-36\right) \eta \right) \rp^4}{36 \eta }.
\end{equation}
Replacing this in the indefinite integral (\ref{eqn:IntegralPotentialHG}), and defining $\Upsilon$ as in (\ref{eqn:Upsilon}) we obtain 
\begin{equation}
\label{eqn:UpsilonHG}
\Upsilon=-\frac{18}{5}\frac{q^4\eta}{16\pi G \ell^2}v=-\frac{6}{5}M.
\end{equation}
The geometric volume defined in (\ref{eqn:GeometricVolume}) can also be calculated using this integral. The result is
\begin{equation}
\label{eqn:GeomThermHG}
V=\frac{3 q^4\eta}{10}v=\frac{12}{5}\Theta. 
\end{equation}

With the knowledge bestowed to us by equation (\ref{eqn:GeometricThermVolume}), we could have arrived to this result without so much trouble. All we need to do is expand the function $\sigma(r)$ around zero. We observe that for small $r$, $\sigma(r)\sim r^{2/3}$. We then identify $\alpha=-2/3$ so, replacing this value in formula (\ref{eqn:GeometricThermVolume}) we get
\begin{equation}
\label{eqn:GeomThermHG2}
\Theta=5/12 V,
\end{equation}
as required.  In general, it is hard to integrate the hypergeometric function, as it will not always reduce to an algebraic expression like above. However, with our result (\ref{eqn:GeometricThermVolume}) we can assert that for general values of the metric parameters
\begin{equation}
\label{eqn:GeomThermHG3}
\Theta=\frac{1}{2}\left(1-\frac{\nu^2 k_{0}^2}{2(D-1)(D-2)}\right)V.
\end{equation}

\subsection{Double branch solution}

The second example, which we call the double branch solution, is a somewhat better known solution. It was first proposed in the 5 dimensional case in \cite{Acena:2012mr} and, more recently, it was generalized to an arbitrary number of spacetime dimensions in \cite{Acena:2013jya}. The chemistry of spherical black hole solutions analogous to this one has also been recently developed in \cite{Astefanesei:2019ehu} and \cite{Rojas:2019hze}. This solution is interesting because it allows to calculate explicitly the beta functions of the relevant deformation couplings in the holographic RG flow, as opposed to the previous example which is dual to a vev-driven flow.

However, for our purposes, this solution is not as convenient as the hypergeometric one. This is due to the fact this solution is not expressed in Poincar\'e coordinates. We can directly verify the Smarr relation and the first law, whence coordinates do not matter, and then integrate the potential to find that there is indeed a proportionality relation between the thermodynamic and geometric volumes. Under the assumptions we used to derive equation (3), we can locally change coordinates and find the asymptotic behaviour of the metric close to $r=0$. We can then verify that the proportionality constant is given by our prescription, but additional work is needed because we need to recover the Poincar\'e coordinates in the infrared. We must also note that this solution has two branches in which the radial coordinate is defined over different intervals. The branch chosen depends on the sign of the scalar field, which does not change in this solution. 

For simplicity, we will work in the 5 dimensional case. The Smarr relation and first law in the D-dimensional case can be derived in a similar fashion, but integrating the potential becomes easier in 5 dimensions.  A dimensionful parameter $\eta$ is introduced to obtain a dimensionless radial coordinate $x$, so the metric takes the form
\begin{equation}
\label{eqn:ConformalAnsatz}
ds^2=\Omega(x)\left(-f(x)dt^2+\frac{\eta^2 dx^2}{f(x)}+\delta_{ij}dx^{i}dx^{j}\right). 
\end{equation} 
Then, with the potential
\begin{align}
16 \pi G V(\phi)&=2\Lambda\left(\frac{(9\nu^2-5)e^{-\phi l_{\nu}}}{8\nu^2}\right)\left(1-\frac{8\mu}{(\nu^2-25)(9\nu^2-25)}\right)\nonumber \\
&\times \left(\frac{(\nu -1) e^ {-\nu  \phi l_{\nu} }}{2 (3 \nu +5)}+\frac{(\nu +1) e^{\nu  \phi l_{\nu} }}{2 (3 \nu -5)}+\frac{5 \left(\nu ^2-1\right)}{9 \nu ^2-25}\right)\nonumber\\
&+2\Lambda \mu   \frac{e^{3 l_{\nu}\phi/2}}{4 \nu ^3} \left[\frac{5 \left(\nu ^2-1\right)}{\nu ^2-25} \left(\frac{e^{-l_{\nu}\nu\phi/2}}{3 \nu -5}+\frac{e^{l_{\nu}\nu\phi/2}}{3 \nu +5}\right)\right.\nonumber\\
&\left. +\frac{1}{3} \left(\frac{(\nu +1) e^{-3l_{\nu}\nu\phi/2}}{(\nu -5) (3 \nu -5)}+\frac{(\nu -1)e^{3\nu l_{\nu}\phi/2}}{(\nu +5) (3 \nu +5)}\right)\right] \label{eqn:PotentialDB}
\end{align}
the Einstein field equations are solved by the metric functions
\begin{equation}
\label{eqn:ConformalFactorDB}
\Omega(x)=\frac{\nu^2 x^{\nu-1}}{\eta^2(x^\nu-1)^2},
\end{equation}
\begin{equation}
\label{eqn:MetricDB}
f(x)=-\frac{\Lambda}{6}+\Lambda\mu\left(\frac{4}{3(\nu^2-25)(9\nu^2-25)}+\frac{x^{5/2}}{12\nu^3}\left(-\frac{x^{-\frac{\nu }{2}}}{\nu -5}-\frac{x^{\nu /2}}{\nu +5}+\frac{x^{-\frac{3 \nu }{2}}}{3 (3 \nu -5)}+\frac{x^{\frac{3 \nu }{2}}}{3 (3 \nu +5)}\right)\right),
\end{equation}
and the scalar field 
\begin{equation}
\label{eqn:ScalarDB}
\phi=l_{\nu}^{-1}\log(x); \qquad l_{\nu}^{-1}=\sqrt{\frac{3(\nu^2-1)}{2}}.
\end{equation}
The numerical constant $\nu\geq 1$ parametrizes different solutions; the value $\nu=1$ corresponds to the vacuum solution. Only in this case, the change to Poincar\'e coordinates can be done exactly. The Hamiltonian method can be used to derive the mass and the tensions, and from them this we can identify the thermodynamic volume. We obtain
\begin{equation}
\label{eqn:ChargesDB}
M=\left(\frac{1}{16 \pi G}\right)\frac{\mu}{16 \eta^4 \ell^2}v, \qquad \uptau_{k}=-\left(\frac{1}{16 \pi G}\right)\frac{\mu}{48 \eta^4 L^{k} \ell^2}v, \qquad \Theta=\frac{1}{4}\frac{\mu}{96\eta^4}v.
\end{equation}
Defining $x_{+}$ to be the largest root of $f(x_{+})=0$, the entropy and temperature are found to be
\begin{equation}
\label{eqn:TempEntropDB}
T=\frac{\mu\left\lvert x_{+}^\nu -1\right\rvert^3}{48\pi\eta\nu^3 x_{+}^{\frac{3}{2}(\nu-1)}\ell^2}, \qquad S=\frac{\nu^3 x_{+}^{\frac{3}{2}(\nu-1)}}{4 G \eta^3\left\lvert x_{+}^\nu -1\right\rvert^3}v.
\end{equation}
Both the equality $4M=3TS$ and the full Smarr relation with volume and tension terms can directly be shown to hold with these quantities. Moreover, the extended first law is also satisfied. This must be done with care, because the parameters that are to be varied independently to check the first law are the compactifying lengths, $\Lambda$ and the parameter $\eta$, which will change the scale and hence the position of the horizon when $x_{+}$ is fixed. 

In order to find the integral of the potential between the Poincar\'e and black brane horizons, we note that the radial coordinate ranges between $0$ and $1$, or $1$ and infinity depending on whether the scalar is negative or positive. In both cases, the conformal infinity is located at $x=1$, and the Poincar\'e horizon is at zero or at infinity respectively. We will work with the second case, which corresponds to stable black branes. The integral is complicated but can be computed exactly for any value of $\nu$. We obtain, following steps similar to the hypergeometric case

\begin{equation}
\label{eqn:IntegralPotentialDB}
16\pi G\int d^3 x \int_{\infty}^{x_{+}}dr\sqrt{-g}V(\phi(x))= -\frac{\nu+1}{3\nu+5}\frac{\mu}{4 \eta^4 \ell^2}v,
\end{equation}
so the relation between the geometric and thermodynamic volumes is
\begin{equation}
\label{eqn:GeomThermDB}
\Theta=\frac{1}{8}\left(\frac{3\nu+5}{\nu+1}\right)V. 
\end{equation}

As an alternative to integrating the lengthy expression for the potential, we may simply use formula (\ref{eqn:UpsilonM}). In order to do this, we need find the metric in Poincar\'e coordinates at least locally close to $x=\infty$, which corresponds to $r=0$. Here, we assume $A(r)$ is finite, so we must have that $\Omega(x(r))\sim r^2$. Under this assumption and from the expression (\ref{eqn:ConformalFactorDB}) for $\Omega(x)$, when $x$ goes to infinity we must have
\begin{equation}
\label{eqn:PoincareDB}
 x\sim r^{-2/(1+\nu)}\quad  \textrm{and} \quad dx^2\sim r^{-2(3+\nu)/(1+\nu)}dr^2. 
 \end{equation}
Using these, and the form of the ansatz (\ref{eqn:MetricAnsatz}), we must have that 
\begin{equation}
\label{eqn:PoincareDB2}
\sigma\sim r^{-\alpha}, \qquad \alpha=\frac{1-\nu}{1+\nu}.
\end{equation} 
From the value of $\alpha$ and using (\ref{eqn:GeometricThermVolume}), we again obtain relation (\ref{eqn:GeomThermDB}). 

\section{Discussion}

In this work we have obtained a very general Smarr relation and First Law of thermodynamics for the gravitational energy of hairy black branes using purely geometric methods. The result for the Smarr relation matches the one obtained from scaling arguments and allows us to define a thermodynamic volume, which can be easily computed from the mass. This volume is the conjugate variable to the pressure in the extended first law, and can be calculated simply from the mass and the cosmological constant. For non-hairy black branes, the thermodynamic and geometric volume are proportional to each other, and this also happens for the general case. However, the constant of proportionality is dependent on the infrared behaviour of the solution, namely
\begin{equation}
\Theta=\frac{1}{2}\left(1+\frac{\alpha}{D-1}\right)V,
\end{equation}
where $\alpha$ characterizes the decay of the $g_{rr}$ component as $r\rightarrow 0$. With this result, we have provided evidence of the conjecture that the integral of the potential behind the horizon is proportional to the thermodynamic volume, if only for this particular horizon topology. To provide a full proof, we must also deal with spherical and hyperbolic horizons. The evidence provided in \cite{Cvetic:2010jb} suggests that a similar result holds.

Let us observe that the integral of the potential turns out to be proportional to the mass, where the constant of proportionality is given by the behaviour of the metric close to the Poincar\'e horizon. Thus, by just knowing the potential inside the horizon, the asymptotic value of the potential and the metric at the infrared, we can know the mass. This means that this information is enough to fix the asymptotics of the blackening factor with a simple prescription. Hopefully this insight will help to find more explicit hairy black brane solutions.
 
A possible extension of this work is to allow for other asymptotic behaviour of the solution. The assumption that the spacetime is asymptotically AdS was relaxed in \cite{Pedraza:2018eey, Kastor:2018cqc}, where similar Smarr relations and first laws hold. The volume gains some terms that are proportional to the trace of the holographic stress-energy tensor. The examples therein holds only for some particular form of the potential, but it is plausible that this result is much more general. With the methods that we have derived, it is likely that such a first law can be generalized. Another possible extension of this work is to include variations of the asymptotic behaviour of scalar field, thus obtaining a full thermodynamics of this spacetime. We would expect that the volume term is modified, as the total energy would include contributions from the scalar field. Deriving such a law could hopefully help us understand more about the holographic interpretation of the volume term, which is still subject to much debate \cite{Johnson:2014yja, Dolan:2014cja, Kastor:2014dra, Karch:2015rpa, Couch:2016exn,Caceres:2015vsa}.

\section*{Acknowledgements}
 The author would like to thank David Kubiz\v n\' ak for the helpful discussions and comments and Robert Myers for inspiring the study of black branes with scalar hair. The work was supported by the Perimeter Institute for Theoretical Physics and by the Natural Sciences and Engineering Research Council of Canada. Research at Perimeter Institute is supported in part by the Government of Canada through the Department of Innovation, Science and Economic Development Canada and by the Province of Ontario through the Ministry of Economic Development, Job Creation and Trade.

\bibliography{references}
\bibliographystyle{JHEP}

\end{document}